# Electronic nature of chiral charge order in the kagome superconductor CsV$_3$Sb$_5$

**Authors:** Zhiwei Wang[1,14*], Yu-Xiao Jiang[2*], Jia-Xin Yin[2*]†, Yongkai Li[1,14*], Guan-Yong Wang[3], Hai-Li Huang[3], Shen Shao[4,5], Jinjin Liu[1,14], Peng Zhu[1,14], Nana Shumiya[2], Md Shafayat Hossain[2], Hongxiong Liu[6,7], Youguo Shi[6,7], Junxi Duan[1,14], Xiang Li[1,14], Guoqing Chang[8], Pengcheng Dai[9], Zijin Ye[10], Gang Xu[10], Yanchao Wang[4,5], Hao Zheng[3], Jinfeng Jia[3], M. Zahid Hasan[2,11,12,13]†, Yugui Yao[1,14]†

**Affiliations:**

[1]Centre for Quantum Physics, Key Laboratory of Advanced Optoelectronic Quantum Architecture and Measurement (MOE), School of Physics, Beijing Institute of Technology, Beijing 100081, China.

[2]Laboratory for Topological Quantum Matter and Advanced Spectroscopy (B7), Department of Physics, Princeton University, Princeton, New Jersey, USA.

[3]School of Physics and Astronomy, Key Laboratory of Artificial Structures and Quantum Control (Ministry of Education), Shenyang National Laboratory for Materials Science, Tsung-Dao Lee Institute, Shanghai Jiao Tong University, Shanghai 200240, China.

[4]State Key Laboratory of Superhard Materials, College of Physics, Jilin University, Changchun 130012, China.

[5]International Center for Computational Method and Software, College of Physics, Jilin University, Changchun 130012, China

[6]Beijing National Laboratory for Condensed Matter Physics and Institute of Physics, Chinese Academy of Sciences, Beijing 100190, China.

[7]University of Chinese Academy of Sciences, Beijing 100049, China.

[8]Division of Physics and Applied Physics, School of Physical and Mathematical Sciences, Nanyang Technological University, Singapore.

[9]Department of Physics and Astronomy, Rice University, Houston, Texas 77005, USA.

[10]Wuhan National High Magnetic Field Center, Huazhong University of Science and Technology, Wuhan, Hubei 430074, China.

[11]Princeton Institute for the Science and Technology of Materials, Princeton University, Princeton, New Jersey 08544, USA.

[12]Materials Science Division, Lawrence Berkeley National Laboratory, Berkeley, California 94720, USA.

[13]Quantum Science Center, Oak Ridge, Tennessee 37831, USA.

[14]Beijing Key Lab of Nanophotonics and Ultrafine Optoelectronic Systems, Beijing Institute of Technology, Beijing 100081, China.

†Corresponding authors, E-mail: jiaxiny@princeton.edu; mzhasan@princeton.edu; ygyao@bit.edu.cn

*These authors contributed equally to this work.

**Kagome superconductors with $T_C$ up to 7K have been discovered over 40 years. Recently, unconventional chiral charge order has been reported in kagome superconductor $KV_3Sb_5$, with an ordering temperature of one order of magnitude higher than the $T_C$. However, the chirality of the charge order has not been reported in the cousin kagome superconductor $CsV_3Sb_5$, and the electronic nature of the chirality remains elusive. In this letter, we report the observation of electronic chiral charge order in $CsV_3Sb_5$ via scanning tunneling microscopy (STM). We observe a 2×2 charge modulation and a 1×4 superlattice in both topographic data and tunneling spectroscopy. 2×2 charge modulation is highly anticipated as a charge order by fundamental kagome lattice models at van Hove filling, and is shown to exhibit intrinsic chirality. We find that the 1×4 superlattices forms various small domain walls, and can be a surface effect as supported by our first-principles calculations. Crucially, we find that the amplitude of the energy gap opened by the charge order exhibits real space modulations, and features 2×2 wave vectors with chirality, highlighting the electronic nature of the chiral charge order. STM study at 0.4K reveals a superconducting energy gap with a gap size 2Δ=0.85meV, which estimates a moderate superconductivity coupling strength with $2\Delta/k_BT_C$=3.9. When further applying a c-axis magnetic field, vortex core bound states are observed within this gap, indicative of clean-limit superconductivity.**

Owing to the unusual lattice geometry, spins on kagome lattices can be geometrically frustrated and form a quantum spin liquid [1,2]. Similarly, a fundamental tight-binding model of kagome lattice considering the nearest neighboring hoping will produce an intriguing electronic band structure containing Dirac cones, flat band, and van Hove singularities. Thus, materials containing kagome lattices are an exciting platform to explore the quantum level interplay between geometry, correlation, and topology. For instance, certain kagome magnets are found to exhibit electronic nematicity, giant spin-orbit tunability, and topological Chern quantum phases [3-16]. Kagome superconductors with competing orders have been identified for over 40 years [17-19], such as $LaRu_3Si_2$ with $T_C$ of 7K and a fundamental kagome band structure [20]. Recently, another layered kagome superconductor $AV_3Sb_5$ (A = K, Rb, Cs) was discovered with $T_C$ up to 2.5K for A=Cs [21-23], providing new research avenues, particularly for transport and STM studies [21-33].

High-quality single crystals of $CsV_3Sb_5$ were grown by a self-flux method with Cs-Sb binary eutectic mixture ($Cs_{0.4}Sb_{0.6}$) as the flux [21]. We found that the molar ratio of $Cs_{0.4}Sb_{0.6}$: $CsV_3Sb_5$ = 15:1 is the best condition for obtaining large crystals with high quality. In order to obtain high-quality crystals, it is necessary to perform surface treatment of raw materials in a hydrogen atmosphere [34,35]. Our crystals have shiny surfaces with a hexagonal shape, and the typical size of crystals is about 3×4×0.5 $mm^3$, as shown in the left inset of Figure 1(b). The crystal structure was analyzed via X-ray diffraction using a θ-2θ scan. The magnetic susceptibility and magnetization were measured in a physical property measurement system magnetometer in vibrating-sample measurement mode. The transport property measurements were carried out in a five-contact configuration so that the longitudinal and Hall resistivity can be taken simultaneously. STM data were obtained with a Pt/Ir tip at 4.2K and 0.4K. The topographic data were obtained in a constant current mode with bias voltage V=50mV and current I=0.05nA. Spectroscopic data were obtained with V=50mV and I=0.5nA V=50mV at 4.2K, and V=5mV and I=0.1nA at 0.4K.

$CsV_3Sb_5$ crystallizes in the hexagonal P6/mmm space group, consists of V-Sb slabs that are intercalated by cesium cation, as illustrated in Fig. 1(a). The most notable feature in this compound is that the vanadium sublattice forms a 2D kagome net. X-ray diffraction pattern displays sharp (00l) reflection peaks, as shown in Fig. 1(b). The full width at half maximum (FWHM) of (004) peak is only 0.07°, as shown in the right inset of Fig. 1(b). Both resistivity data [Fig. 1(c)] and susceptibility data [Fig. 1(d)] show two phase transitions. The 90K transition corresponds to the development of a charge order, and the 2.5K transition corresponds to the development of superconductivity. Below the charge order transition temperature, previous studies reveal the possibility of anomalous Hall effect [24,25]. Similar to these studies, at 5K, both the magnetization data [Fig. 1(f)] and the Hall resistivity data [Fig. 1(e)] as a function of the magnetic field applied along the c-axis show a double kink feature. By further subtracting a linear background as the ordinary Hall response, this unusual Hall response would resemble an anomalous Hall effect, as shown in the inset of Fig. 1(e). Our low-field magnetization loop data in the inset of Fig. 1(f) also confirms the absence of magnetic hysteresis. The observation of anomalous Hall effect and increasing magnetization below charge ordering temperature with decreasing temperature implies the highly unconventional nature of the charge order.

In this material, the V and Sb layers have a stronger chemical bonding, and the material tends to cleave between Cs and Sb layers. The Sb surface is most interesting, as it is strongly bonded to the V kagome lattice. Previous STM studies have resolved Sb honeycomb surfaces in $AV_3Sb_5$, and studied the charge order and surface superlattices [20-26]. We study $CsV_3Sb_5$ with STM at 4.2K. Through cryogenic cleaving, we have also obtained large Sb surfaces in $CsV_3Sb_5$, as shown in Fig. 2(a). Similar to $KV_3Sb_5$ [28], we observe a soft charge order gap opening from -23meV to +35meV [Fig. 2(b)]. The Fourier transform of the topography confirms the 2×2 charge order as marked by the shaded red region in Fig. 2(c), which is ubiquitous for all $AV_3Sb_5$ compounds. In defect-free areas of $KV_3Sb_5$ and $RbV_3Sb_5$, the 2×2 charge order is found to display a chirality [28,29], where its three vector peaks have different intensities. We thus perform a spectroscopic map in a defect-free area [Fig. 1(d)] for $CsV_3Sb_5$ to check the existence of the chirality. The energy-resolved intensity of three 2×2 vector peaks is displayed in Fig. 2(e), where we observe clear intensity differences. The observed anisotropy can be due to a chiral charge order as observed and discussed in certain transition-metal dichalcogenides and high-temperature superconductors [36,37]. The chirality can be defined as the counting direction (clockwise or anticlockwise) from the lowest to the highest pair of vector peaks. By contrast, near a defect-rich region [Fig. 2(f)], we find that there is almost no chirality for the charge order [Fig. 2(g)], which suggests the existence of extrinsic effects. In both cases, the intensity of $Q_1$ is strongest, and along this direction, there are additional 1×4 superlattice vector peaks in the q-space data [Fig. 2(c)]. This observation suggests an interplay between the 1×4 superlattice and 2×2 charge order.

The 1×4 superlattice is not observed on the Sb surface in $KV_3Sb_5$ [28], certain Sb surface in $CsV_3Sb_5$ [30], and the K/Cs/Rb surfaces in $AV_3Sb_5$. We also find many domains of the 1×4 superlattice, including 120-degree domains [Fig. 3(a)] and phase shift domains [Fig. 3(b)]. The domain size can be very small (less than 10nm), which will not be energy favorable if it is bulk in nature. In addition, bulk X-ray measurements have detected the 2×2 charge order but have not yet observed the 1×4 superlattice [21,28]. Therefore, these pieces of experimental evidence taken together suggest that the 1×4 superlattice effect is likely to be fragile and of surface origin. For another example, it is known that the 1×2 superlattice observed by STM in $BaFe_2As_2$ is a polar surface effect [38], and there can be various small domain walls of the 1×2 superlattices, as similarly observed here. To explore the surface origin possibility, we perform first-principles calculations. The geometry optimizations of surface atoms were performed in the framework of density-functional theory with the projector augmented wave method and the PBE exchange-correlation functional [39], which are implemented in the Vienna Ab into Simulation Package [40]. We built two types of 1×4 supercells slabs, repeating the primitive cell in two different directions: parallel (Type-A) and vertical (Type-B) to the a-axis. After fully relaxed surface atoms, we found only the Type-B slab is stabilized into the 1×4 supercell, which is consistent with our experimental measurements. Our calculations show that 1×4 supercells are due to the in-plane shifting of V atoms relative to surface Sb atoms, as indicated by the black arrows in Fig. 3(c).

Previous first-principles calculation [28,41] in $AV_3Sb_5$ has also confirmed that the 2×2 charge order is a bulk effect and is consistent with a hexamer and trimer formation [28] (or termed as inverse star of David formation [41]) in the V kagome lattice. The chirality feature of the charge order, however, is beyond the current first-principles calculation. Prior to any theoretical discussion of the origin of chirality, it is crucial to explore whether the chirality is of electronic nature. An issue in the differential conductance mapping presented in Fig. 2 is that it cannot distinguish the electronic contribution from the structural contribution, as the tunneling junction setup can often have a dependence on structure features. However, it is generally believed that the energy gap of the charge order is of electronic origin. We therefore systematically measure the energy gap of the charge order at each spatial point and map out the spatial distribution of the charge order gap value as shown in Fig. 4(a), which reveals a striking spatial modulation of the energy gap value. The Fourier transform of the gap map in Fig. 4(b) shows the 2×2 vector peaks with clear anisotropy, thus similarly defining a chirality in the electronic charge order. We have further checked this phenomenon in the cousin kagome superconductor $RbV_3Sb_5$ in Figs. 4(c) and (d), and observed similar chirality of the electronic charge order. In Fig. 4 (e), we show typical linecut dI/dV spectra along the line draw in Fig. 4(c), further illustrating the charge order gap modulation in real-space.

The observation of electronic chiral charge order is unprecedented in the studying of quantum materials, and can be highly related to the electronic nature of kagome lattice at van Hove filling as in $AV_3Sb_5$ [42-46]. In contrast to the honeycomb lattice, the kagome lattice has three sublattices, and the kagome lattice at van Hove filling exhibits unequal predominant sublattice occupancy [42]. With extended Coulomb interactions considered, this sublattice interference mechanism will produce a chiral charge order with relative angular momentum [44]. Phenomenologically, a chiral charge order parameter has also been proposed to interact with the underlying topological band structure of this material to produce a large intrinsic anomalous Hall effect [28,43,44,45], which is consistent with the transport data. Another anomalous feature of the chiral charge order is that it can carry orbital currents [28,43,44,46], analogous to the fundamental Haldane model [47] and Varma model [48]. A tantalizing visualization of the orbital magnetism is recently reported by muon spin spectroscopy, and a spontaneous internal magnetic field is observed just below the charge ordering temperature [49].

Finally, we explore the superconducting tunneling spectrum at 0.4K, well below the superconducting transition temperature. Consistent with previous reports [30,32,33], the tunneling spectrums show an energy gap with a gap size $2\Delta=0.85meV$ determined by the coherence peak-to-peak distance [Fig. 5(a)(b)]. Accordingly, we obtain the superconductivity coupling strength $2\Delta/k_BT_C=3.9$, which is slightly larger than the standard BCS value and suggests the electron pairing is close to a moderate coupling. By applying a 50mT magnetic field along the c-axis, we observe a zero-energy vortex core state [Fig. 5(c)(d)]. The formation of a clear vortex core bound state implies that the superconducting state is in the clean-limit, where the electron mean free path is much longer than the superconducting coherent length. One notable observation in the superconducting state is that the gap is shallow, and the measured gap is filled with substantial states. The origin of the in-gap states deserves future attention with lower temperature and higher energy resolution measurements. If such states are intrinsic and robust, then it is worthy of discussing further their relationship with the tantalizing broken-symmetry [32][49] of the superconducting state.

**Acknowledgement**
The work at Beijing Institute of Technology was supported by the National Key R&D Program of China (Grant No. 2020YFA0308800), the Natural Science Foundation of China (Grant Numbers. 92065109, 11734003, 12061131002), the Beijing Natural Science Foundation (Grant No. Z190006), and the Beijing Institute of Technology (BIT) Research Fund Program for Young Scholars (Grant No. 3180012222011). Z.W. thanks the Analysis & Testing Center at BIT for assistance in facility support. Experimental and theoretical work at Princeton University was supported by the Gordon and Betty Moore Foundation (GBMF4547; M.Z.H.). Y.S. was supported by the National Natural Science Foundation of China (U2032204), and the K. C. Wong Education Foundation (GJTD-2018-01). G.C. would like to acknowledge the support of the National Research Foundation, Singapore under its NRF Fellowship Award (NRF-NRFF13-2021-0010) and the Nanyang Assistant Professorship grant from Nanyang Technological University. P. D. is supported the US Department of Energy (DOE), Basic Energy Sciences (BES), under Contract No. DE-SC0012311. S.S. and Y.W was supported by the National Natural Science Foundation of China under Grants No. 11822404 and No. 11774127.

**Figures:**

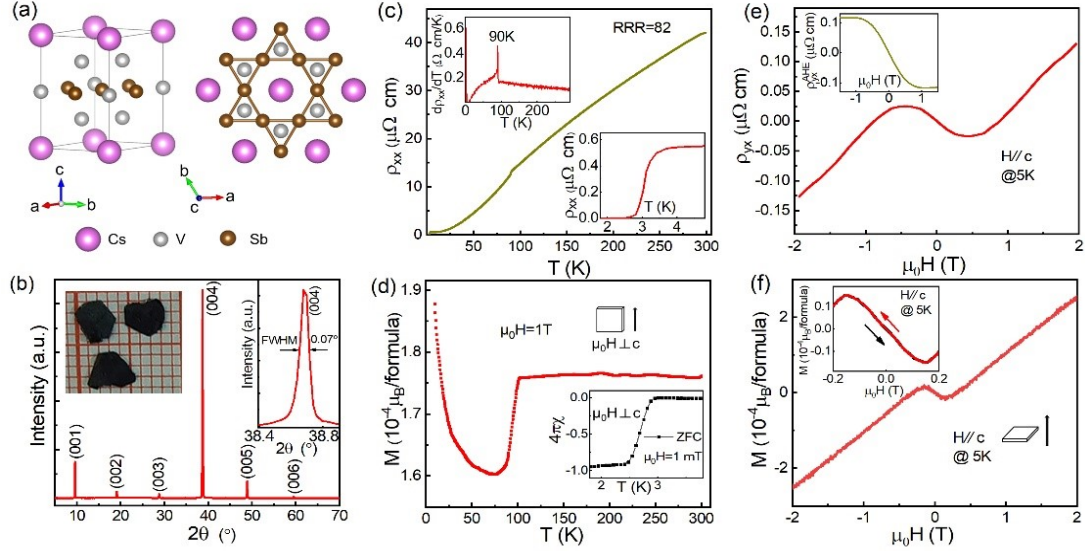

Figure 1. (a) Crystal structure of $CsV_3Sb_5$. (b) X-ray diffraction pattern of as-grown single crystal, showing sharp (00l) reflections. The left inset shows the crystal image, and the right inset shows an enlarged view of the (004) peak. (c) Temperature dependence of $\rho_{xx}$ for $CsV_3Sb_5$. The top inset shows $d\rho_{xx}/dT$ as a function of temperature, and the bottom inset shows the $\rho_{xx}(T)$ behavior near the superconducting transition. (d) Temperature dependence of the magnetic susceptibility in $CsV_3Sb_5$ under 1T for zero-field cooling measurement with $H\perp c$, and the inset shows diamagnetic behavior under 1mT with $H\perp c$ near the superconducting transition. (e) and (f) Hall resistivity $\rho_{yx}$ and magnetization $M$ as a function of the magnetic field measured at 5 K with H//c, respectively. The inset of (e) shows extracted anomalous Hall signal by subtracting the local linear ordinary Hall background. The inset of (f) shows the zoom-in magnetization data, indicative of the absence of magnetic hysteresis.

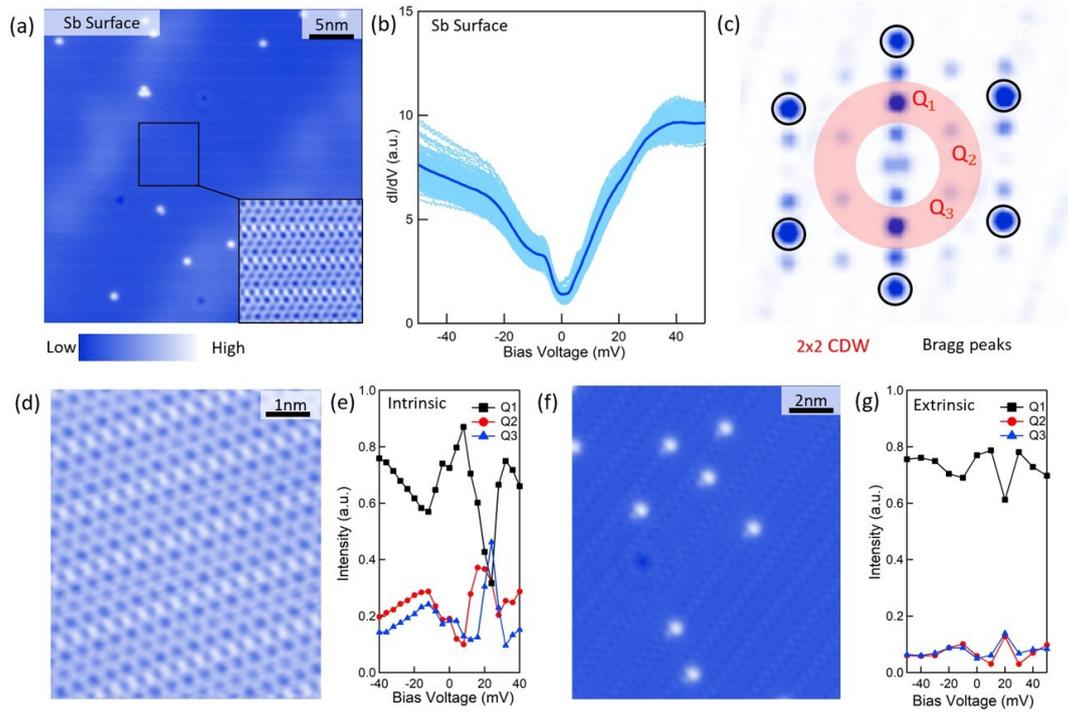

Figure 2. (a) Topographic image of the Sb surface. The inset shows a zoom-in topography of the Sb honeycomb lattice. (b) The dark blue curve is the spatially averaged dI/dV spectrums (light blue curves) for a 20nm×20nm Sb surface. (c) Fourier transform of an Sb topographic image, showing 2×2 charge order vector peaks (red shaded area), and 1×4 vector peaks along the $Q_1$ direction. (d)(e) A defect-free Sb surface and intensity of charge ordering peaks at different bias voltages. The three 2×2 charge order vector peaks have different intensities, defining an intrinsic chirality. (f)(g) A defect-rich Sb surface and intensity of charge ordering peaks at different bias voltages. This area shows almost no chirality, which can be an extrinsic effect.

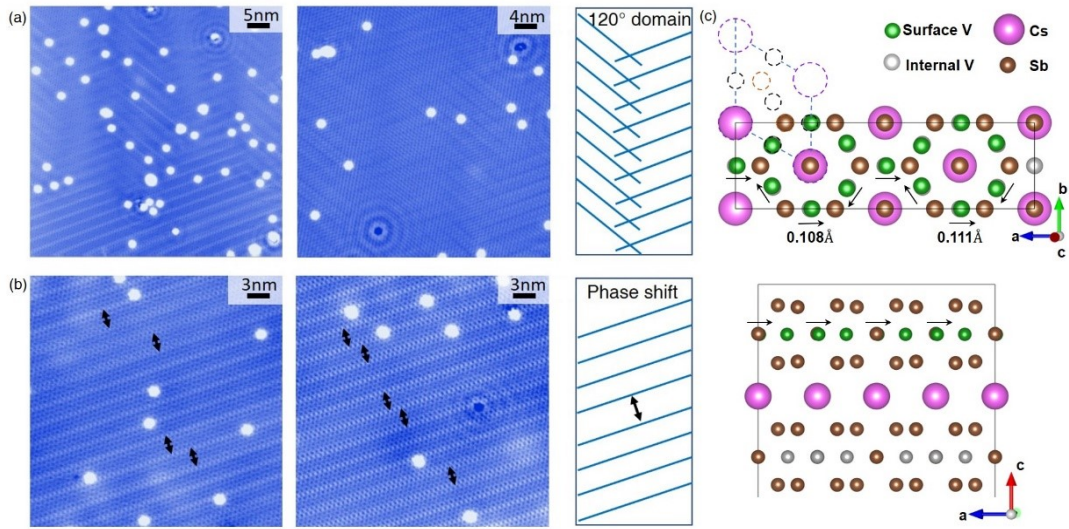

Figure 3. (a) Topographic (left two) and schematic (right) images showing 120 degree domains for the 1×4 superlattices. On the right is the schematic image for the 120 degree domains. (b) Topographic (left two) and schematic (right) images showing phase shift domains for the 1×4 superlattices. (c) First-principles calculation of the Sb surface, indicating the surface origin of the 1×4 superlattices. The green atoms represent surface V atoms. Arrows show the in-plane movements of surface V atoms. The rhombic cell in the dashed line is the primitive cell. The thickness of the slab is 29.719 Å. The enthalpy of the relaxed type-B supercell slab is ~7 meV/f.u. lower after surface optimizations at 0 K, indicating that the surface reconstruction occurs spontaneously. Kinetic cutoff energy of 600 eV and spacing of $2\pi \times 0.03$ Å$^{-1}$ for Monkhorst-Pack k-mesh sampling were adopted to give well converged total energies (~1 meV/atom).

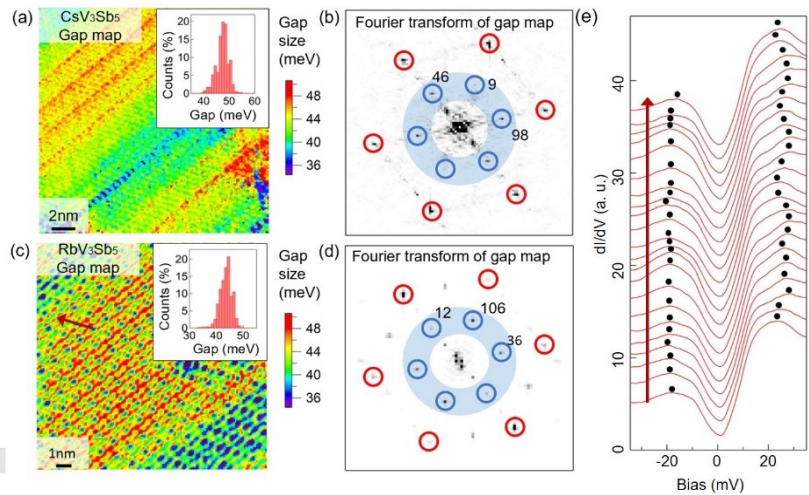

Figure 4. (a) Energy gap map in CsV$_3$Sb$_5$ showing the spatial distribution of the amplitude of the energy gap opened by the charge order. The inset shows the histogram of the gap distribution. (b) Fourier transform of the gap map in (a). The 2×2 vector peaks are marked with blue circles, and their intensities are noted respectively. (c) Gap map for RbV$_3$Sb$_5$. (d) Fourier transform of the gap map in (c). (d) dI/dV data along the line marked by the arrow in (c), showing the modulation of the charge order gap. The black dots mark the peaks on both the positive and negative sides of the spectrum, between which defines the gap size.

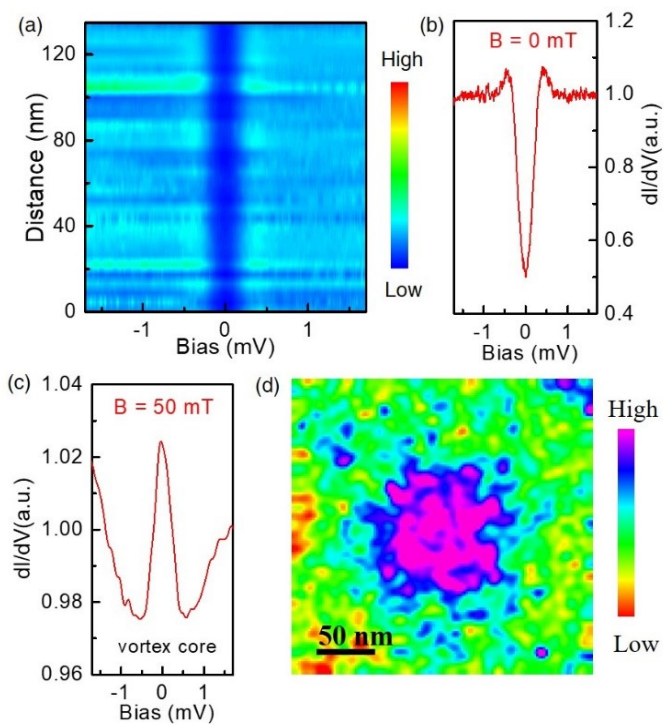

Figure. 5 (a) Intensity plot of line-cut spectrums taken at T = 0.4K, showing superconducting gaps. (b) A typical superconducting gap spectrum. (c) Vortex core spectrum showing a zero-energy state. (d) A vortex core map taken at zero-energy.